\documentclass[aps,pra,twocolumn,showpacs]{revtex4-1}

\usepackage{graphicx}

\begin{document}

\title{Direct observation of temporal coherence\\ by weak projective measurements of photon arrival time}

\author{Holger F. Hofmann}
\email{hofmann@hiroshima-u.ac.jp}
\author{Changliang Ren}
\affiliation{
Graduate School of Advanced Sciences of Matter, Hiroshima University,
Kagamiyama 1-3-1, Higashi Hiroshima 739-8530, Japan}
\affiliation{JST, CREST, Sanbancho 5, Chiyoda-ku, Tokyo 102-0075, Japan
}

\begin{abstract}
We show that a weak projective measurement of photon arrival time can be realized by controllable two photon interferences with photons from short-time reference pulses at a polarization beam splitter. The weak value of the projector on the arrival time defined by the reference pulse can be obtained from the coincidence rates conditioned by a specific output measurement. If the weak measurement is followed by a measurement of frequency, the coincidence counts reveal the complete temporal coherence of the single photon wavefunction. Significantly, the weak values of the input state can also be obtained at higher measurement strengths, so that correlations between weak measurements on separate photons can be observed and evaluated without difficulty. The method can thus be used to directly observe the non-classical statistics of time-energy entangled photons.
\end{abstract}


\maketitle
\section{Introduction}

Photons are among the most accessible and well-controlled quantum systems available. The high level of coherence achieved by optical technologies makes it possible to observe a wide variety of quantum effects using spatial interference and polarization. It is also possible to use the temporal coherence of photon states to encode quantum information and to observe non-classical effects in time \cite{Tit00,Zbi01,Mar02,Mar03,Rie04,Cab09,Val10,Fed11}. However, the time-energy degree of freedom is more difficult to access than polarization or spatial coherence, since linear optics cannot convert the frequency of photons, limiting the unitary operations that can be performed with passive linear elements to those that conserve photon energy. Moreover, time resolved detection is limited by the temporal resolution of detector systems, so that the coherence of broadband entanglement is difficult to measure and verify experimentally \cite{Har07,Nas08,Don09,Svo09,Sen10}. 

To overcome the low time resolution associated with direct photon detection, it is interesting to consider measurement strategies that make use of quantum interactions between the signal photon and a short-time probe. If a sufficiently strong optical non-linearity can be achieved, the probe can be a classical field, e.g. in an upconversion process \cite{Kuz08a,Kuz08b}. If only passive linear optics elements are used, the possible interactions between the signal photon and a short-time probe photon are limited to two-photon interference effects. Linear optics measurements of photon arrival time must therefore be based on the bunching effect caused by two-photon interference of a signal photon and a photon from a short-time reference pulse at a beam splitter \cite{Was07}. As shown in previous investigations \cite{Ren11}, time information can be obtained directly using single-time reference pulses since bunching indicates coincidence between the signal photon and the reference pulse. However, frequency information can only be obtained indirectly, by analyzing the bunching effects of reference pulses in temporal superpositions \cite{Was07,Ren11}. 

In this paper, we consider the possibility of measuring time and frequency information simultaneously by performing a frequency resolved measurement on the output state following a variable strength two-photon interference. Effectively, the photon bunching effect is used to realize a weak measurement of photon arrival time, where the coincidence rate between signal and probe photon corresponds to the weak value of the projector on the arrival time interval defined by the probe \cite{Aha88,Rus02}. Importantly, this is somewhat different from previously discussed weak measurements of photon arrival time \cite{Ahn04,Wan06}, where the count rates corresponded to the weak value of arrival time, and not to the weak value of the projector on an arrival time interval. Specifically, the weak measurement discussed in the following corresponds to the weak projective measurements of position used to determine the wavefunction of a particle in \cite{Lun11}. The weak projective measurement of photon arrival time therefore provides a particularly direct measurement of temporal coherence in single photon states. Moreover, the complete density matrix of the photon can be described in terms of a complex valued joint probability of time and frequency, as described in \cite{Lun12}. 

The experimental realization of the weak measurement can be achieved by using photon polarization as an auxiliary degree of freedom. The principle used to control the measurement interaction is then similar to the one used in the realization of a weak measurement of polarization with a linear optics phase gate \cite{Pry04,Pry05,Ral06}, where the input and output polarizations of the probe photon are varied to obtain the desired measurement strength. If a polarization beam splitter is used to realize the two-photon interference, the interaction strength will depend on the particular combination of input and output polarizations, independent of the temporal quantum states of the signal photon and the probe photon. When ultrashort optical pulses are used to define the state of the probe photon $\mid t \rangle$, the polarization dependent bunching effect is described a weak projection $\mid t \rangle\langle t \mid$, which approximates the projection onto an eigenstate of time at a time resolution given by the pulse length. Since the interaction is weak, the transmitted signal photon still carries much of the original frequency information of the input state, and this information can be accessed by a subsequent frequency resolved photon detection. 

In the following, we first introduce the measurement procedure for the weak projective measurement of photon arrival time, including all measurement errors caused by the non-vanishing interaction at the polarizing beam splitter. This analysis shows that the quantum interference effect responsible for the weak projection is independent of the actual measurement strength. It is therefore possible to obtain the same result both in the extreme weak limit, and at intermediate strengths. We then show how the wavefunctions in time and in frequency can be obtained from the measurement data. Significantly, the results are observed directly in the data if the polarization independent background is subtracted. It is therefore straightforward to apply the approach to multi-photon wavefunctions, making the non-classical features of entanglement accessible to direct experimental investigation. 

\section{Weak projective measurement of spatiotemporal photon states}

\begin{figure}[th]
\begin{picture}(240,180)
\put(0,0){\makebox(240,180){\vspace{-2cm}\hspace{-0.8cm}
\scalebox{0.5}[0.5]{
\includegraphics{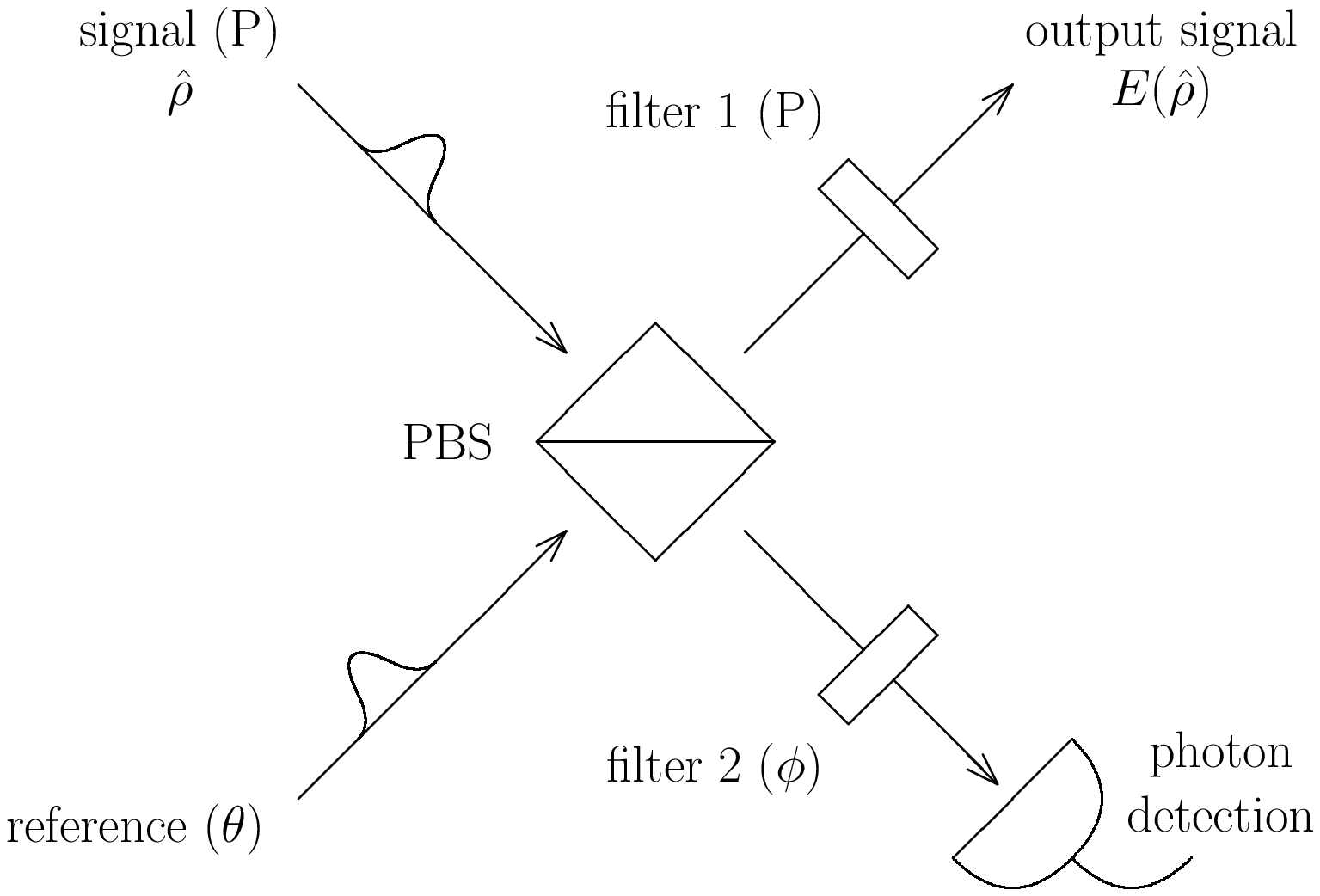}}}}
\end{picture}
\caption{\label{fig1} Illustration of the experimental setup for a weak measurement of time using a short time reference pulse and a polarizing beam splitter (PBS) that transmits horizontally (H) polarized photons and reflects vertically (V) polarized photons. P denotes a positive superposition of H and V, $\theta$ indicates a linear polarization rotated by an angle of $\theta$ from the horizontal, and $\phi$ indicates the ellipticity of a polarization aligned along the diagonal (P-polarization for $\phi=0$). The rate of photon detection after filter 2 determines the measurement result, and $E(\hat{\rho})$ describes the effects of the measurement interaction on the transmitted signal state.}
\end{figure}

We consider a simple experimental setup based on a single polarizing beam splitter that transmits horizontal (H) polarization and reflects vertical (V) polarization. If the signal photon and a reference photon are injected into the beam splitter from opposite sides, and only the detection of photons exiting from opposite sides is considered, the two-photon interference between the transmission and the reflection of both photons can be controlled by the choice of polarization in the input and the output of the beam splitter. Note that the reference photon can be obtained from a suitably weakened coherent source if the statistical contributions of multi-photon inputs from the reference can be neglected \cite{Ren11}. In principle, the photon polarization serves as an additional degree of freedom in a quantum circuit, where the polarization beam splitter represents a quantum controlled-NOT interacting the path and the polarization. Since the polarization is merely an ancillary system, the specific implementation can be varied. For simplicity, we choose the configuration illustrated in Fig. \ref{fig1}, where the polarization of the signal photon is kept constant and the polarizations of reference input and reference detection are modified to achieve the desired effect. Specifically, the signal photon input is polarized along the diagonal, corresponding to the positive (P) superposition of H and V and the interaction strength is varied by rotating the polarization of the reference photon from H (only transmission, no interference) to P (equal superposition of transmission and reflection, maximal interference). The complete input photon states can be described as product states of the time-energy degree of freedom and the polarization,
\begin{eqnarray}
\mid \mbox{signal} \rangle &=& \mid \psi \rangle \otimes \frac{1}{\sqrt{2}}\left(\mid H \rangle + \mid V \rangle \right)
\nonumber
\\
\mid \mbox{reference} \rangle &=& \mid \Phi_{\mathrm{ref.}} \rangle \otimes \left(\cos \theta \mid H \rangle + \sin \theta \mid V \rangle \right),
\end{eqnarray}
where $\mid \psi \rangle$ is the time-energy input state and $\mid \Phi_{\mathrm{ref.}} \rangle$ represents the well-defined pulse shape of the reference photon. 

We now consider only those beam splitter outputs where one photon is found in each output port. Thus, the two photon polarization states in the output can only be $\mid HH \rangle$ if both photons are transmitted, or $\mid VV \rangle$ if both photons are reflected. To obtain interference between these two components, it is necessary to apply polarization filtering to both output ports to eliminate the HV information. The polarizations selected should be equal superpositions of H and V, where the ellipticity of one of the polarizations can be used to select a specific phase $\phi$ for the superposition,
\begin{eqnarray}
\mid \mbox{filter1} \rangle &=& \frac{1}{\sqrt{2}}\left(\mid H \rangle + \mid V \rangle \right)
\nonumber
\\
\mid \mbox{filter2} \rangle &=& \frac{1}{\sqrt{2}}\left(\mid H \rangle + \mathrm{e}^{i \phi} \mid V \rangle \right).
\end{eqnarray}
After the polarization filters are applied, the output state describes a two-photon interference between the temporal input states and the swapped temporal input states controlled in both strength and phase by the polarizations used in the input and the output,
\begin{eqnarray}
\lefteqn{
\mid \mbox{out} \rangle =}
\nonumber \\ && \hspace*{0.4cm}
 \frac{1}{2\sqrt{2}} \left(
\cos \theta \mid \psi; \Phi_{\mathrm{ref.}} \rangle - \mathrm{e}^{-i\phi}\sin \theta \mid \Phi_{\mathrm{ref}}; \psi \rangle\right). 
\end{eqnarray}
For $\theta=\pi/4$ and $\phi=0$, this two-photon interference effect corresponds to the standard photon bunching observed at a beam splitter of reflectivity 1/2, where only the anti-symmetric components of the two-photon wavefunction can contribute to the coincidence counts of photons detected in opposite output ports. However, the interaction strength can now be reduced by choosing a probe polarization closer to $H$, as indicated by a lower value of $\theta$. In this case, the probability of transmission is higher than the probability of reflection, and the output of the first port is still close to the original state of the input photon. In the limit of $\theta \to 0$, the unchanged signal photon state is transmitted to the first port, while the unchanged probe state is transmitted to the second port. Hence, the photon exiting from the first port carries the output signal, while the photon exiting from the second port carries the information associated with the measurement result. 

The meter readout of the weak projection on photon arrival time is given by the count rates for the detection of the probe photon exiting from the second port. If a broadband detector is used, no time or frequency information is obtained in the detection, and the temporal quantum state of the photon in the second port can be traced out. The effect of the probe photon detection on the state of the signal photon exiting from the first port can then be described by a quantum operation on the input density matrix $\hat{\rho}=\mid \psi \rangle\langle \psi \mid$, so that the conditional output state of the probe photon detection is given by
\begin{eqnarray}
\label{eq:process}
\lefteqn{E(\hat{\rho}) = \hat{M}(\theta,\phi) \hat{\rho} \hat{M}^\dagger(\theta,\phi)} 
\nonumber \\ && \hspace*{0.4cm}
+ \frac{(\sin \theta)^2}{8}\left(1-\langle \Phi_{\mathrm{ref.}}\!\mid \hat{\rho} \mid\! \Phi_{\mathrm{ref.}} \rangle\right)
\mid\! \Phi_{\mathrm{ref.}}\rangle\langle \Phi_{\mathrm{ref.}} \!\mid,
\end{eqnarray}
where the essential part of the measurement process is described by the operators
\begin{eqnarray}
\label{eq:operator}
\lefteqn{\hspace{-0.5cm}
\hat{M}(\theta,\phi) =}
\nonumber \\ &&
 \frac{1}{2\sqrt{2}}\cos\theta \left(\hat{I}- \mathrm{e}^{-i \phi}\tan\theta \mid  \Phi_{\mathrm{ref.}}\rangle\langle \Phi_{\mathrm{ref.}}\mid \right).
\end{eqnarray}
Since this operation describes a measurement process, the trace of the output state gives the probability of obtaining the measurement result. In the present context, this means that the probability of detecting the probe photon is given by 
\begin{equation}
\label{eq:counts}
\mbox{Tr}\left(E(\hat{\rho})\right) = \frac{1}{8} - \frac{1}{4}\sin \theta \cos\theta \left(\cos \phi \langle \Phi_{\mathrm{ref.}}\!\mid \hat{\rho} \mid\! \Phi_{\mathrm{ref.}} \rangle \right).
\end{equation}
This result shows that the detection of a probe photon depends on the projective measurement probability of the reference pulse state $\mid \Phi_{\mathrm{ref.}} \rangle$ in the input state $\hat{\rho}$, where the polarizer settings control both the strength of the measurement and the sign of the contribution to the detection probability. 

To determine the difference between the experimental count rate and the background count rate associated with the probability of $1/8$ for orthogonal input and reference states, it may be useful to change the phase $\phi$ that represents the ellipticity of the detected probe photon polarization. Specifically, it is possible to subtract the background by taking the difference between the opposite diagonal polarizations polarizations at $\phi=0$ and at $\phi=\pi$. The experimental result for the weak projective measurement can then be given by the normalized difference of the count rates $C_{\phi=\pi}$ and $C_{\phi=0}$, which are proportional to the probabilities in Eq.(\ref{eq:counts}), so that 
\begin{equation}
\label{eq:expC}
\frac{C_{\phi=\pi}-C_{\phi=0}}{C_{\phi=\pi}+C_{\phi=0}} = 2 \sin\theta \cos\theta  \; \langle \Phi_{\mathrm{ref.}}\!\mid \hat{\rho} \mid\! \Phi_{\mathrm{ref.}} \rangle.
\end{equation}
Thus, the normalized count rate difference between orthogonal polarizations of the probe photon can be used as the output value of the meter in a weak projective measurement of the input photon. 

Since the weak measurement result is obtained from the count rates of the probe photon, the signal photon is available for additional measurements of its time-energy degree of freedom in the output port. It is therefore possible to obtain post-selected measurement results by performing appropriate final measurements $\mid f \rangle$ on the signal photon in the output of the weak measurement setup.

\section{Post-selected weak measurement}

In the weak measurement limit of $\theta \ll 1$, the terms proportional to $(\sin \theta)^2$ in Eq.(\ref{eq:process}) can be neglected and the measurement process is approximately described by the measurement operator $\hat{M}(\theta,\phi)$. As pointed out in \cite{Hof10}, this operator represents a general weak measurement of the projector on the the reference state. It is therefore possible to express the effects of a final measurement on the signal photon in terms of the post-selected weak values of the projection on $\mid \Phi_{\mathrm{ref.}} \rangle$. For an input state $\mid \psi \rangle$, the probability of detecting the probe photon in the second port and the signal photon in a final state $\mid f \rangle$ in the first port can be given by the coincidence count rate
\begin{eqnarray}
\lefteqn{C(\Phi_{\mathrm{ref.}}, f) = |\langle f \mid \hat{M} \mid \psi \rangle|^2}
\nonumber \\ &&
= \frac{1}{8} |\langle f \mid \psi \rangle|^2 \left(1-2 \theta \;
\mbox{Re}\left(e^{-i \phi} \frac{\langle f \mid \Phi_{\mathrm{ref.}}\rangle \langle\Phi_{\mathrm{ref.}} \mid \psi \rangle}{\langle f \mid \psi \rangle} \right)
 \right), \nonumber \\ 
\end{eqnarray}
where all contributions of order $\theta^2$ or higher have been neglected. In the extreme weak limit of $\theta \to 0$, the ratio between this count rate and the count rate of Eq.(\ref{eq:counts}) is given by the probability of post-selection, $|\langle f \mid \psi \rangle|^2$. Weak measurement results are given by contributions linear in the measurement strength $\theta$. The modifications of these first order probabilities by post-selection correspond to a replacement of the expectation value in Eq.(\ref{eq:counts}) with the post-selected weak value of the projector,
\begin{equation}
\frac{\langle f \mid \Phi_{\mathrm{ref.}}\rangle \langle\Phi_{\mathrm{ref.}} \mid \psi \rangle}{\langle f \mid \psi \rangle} = \frac{\langle f \mid \Phi_{\mathrm{ref.}}\rangle \langle\Phi_{\mathrm{ref.}} \mid \hat{\rho} \mid f \rangle}{\langle f \mid \hat{\rho} \mid f \rangle}.
\end{equation}
The real part of this weak value can be evaluated directly from the coincidence count rates of signal photons and probe photons, where the signal photons are detected in the P-polarized output state $\mid f \rangle$ and the probe photons are detected in the diagonal polarizations at $\phi=0$ and at $\phi=\pi$ without any spectral or temporal resolution. Using only the counts obtained for the specific signal output $\mid f \rangle$, the post-selected meter output can be determined by taking the ratio of the count rate differences and the count rate sums for the two probe polarizations, 
\begin{eqnarray}
\label{eq:expost}
\lefteqn{\hspace*{-0.8cm} \frac{C_{\phi=\pi}-C_{\phi=0}}{C_{\phi=\pi}+C_{\phi=0}} =}
\nonumber \\ &&
2 \theta \; \mbox{Re}\left(\frac{\langle f \mid \Phi_{\mathrm{ref.}}\rangle \langle\Phi_{\mathrm{ref.}} \mid \hat{\rho} \mid f \rangle}{\langle f \mid \hat{\rho} \mid f \rangle} \right).
\end{eqnarray}
Thus the meter value obtained for a post-selected weak measurement is equal to the theoretically defined weak value of the projector $\mid \Phi_{\mathrm{ref.}}\rangle \langle\Phi_{\mathrm{ref.}} \mid$ for the initial state $\hat{\rho}$ and the post-selection of $\mid f \rangle$. The imaginary part of this weak value can be obtained in the same setup by changing the settings of the polarizers in the probe output to the circular polarizations defined by $\phi=\pi/2$ and $\phi=3\pi/2$. 

The present setup can be used to realize weak measurements of any kind of spatiotemporal projection operator defined by a corresponding reference pulse shape. The interaction strength of the measurement is given by the angle $\theta$ of the reference polarization and can be varied between the weak limit at $\theta\ll 1$ and a maximally projective measurement at $\theta=\pi/4$. In the weak limit, the output probabilities of $\mid f \rangle$ are unchanged and the effects of measurement back-action can be neglected. However, higher order terms in $\theta$ describe back-action effects that modify the probabilities of $\mid f \rangle$. In the following, we analyze the effects of the complete measurement process given by Eq.(\ref{eq:process}) and show that it is possible to identify the contribution of the finite measurement back-action in the experimental data. It is therefore possible to subtract the events caused by measurement back-action from the experimentally observed count rates, permitting a direct experimental determination of weak values that is also valid in the strong measurement limit \cite{Suz12,Hof13}.

\section{Frequency measurement on the signal output}

Since frequency can be measured with conventional methods, it is of particular interest to realize weak projective measurements of time by using short-time reference pulses with peak time $t$. For sufficiently short pulse times, the reference state $\mid \Phi_{\mathrm{ref.}}\rangle = \mid t \rangle$ approximates an eigenstate of photon arrival time. If the weak measurement of time is followed by a final measurement of photon frequency $\omega$, the coincidence count rates obtained at different times and frequencies are given by 
\begin{eqnarray}
\label{eq:result}
C(\omega,t) &=& \langle \omega \mid E(\hat{\rho}) \mid t \rangle|^2 
\nonumber \\
&=& \frac{1}{8}\cos^2 \theta \langle \omega \mid \hat{\rho} \mid \omega \rangle + 
\frac{1}{8}\sin^2 \theta |\langle \omega \mid t \rangle|^2 
\nonumber \\ &&
- \frac{1}{4} \sin \theta \cos \theta \mathrm{Re}\left( \mathrm{e}^{i \phi}\langle \omega \mid t \rangle \langle t \mid \hat{\rho}\mid \omega \rangle \right)
\nonumber \\ &&
\end{eqnarray}
for all values of measurement strength $\theta$. In this count rate, the contribution of the weak value of $\mid t \rangle\langle t \mid$ is given by  
\begin{equation}
\label{eq:weak}
\frac{\langle \omega \mid (\mid t \rangle\langle t \mid)\mid \psi \rangle}{\langle \omega \mid \psi \rangle} = \frac{\langle \omega \mid t \rangle \langle t \mid \hat{\rho}\mid \omega \rangle}{ \langle \omega \mid \hat{\rho} \mid \omega \rangle}. 
\end{equation}
In the weak limit, this ratio can be obtained from the ratio of the difference and the sum of the count rates observed for orthogonal probe photon polarizations, as described by Eq. (\ref{eq:expost}). This procedure can be applied because the back-action effects of the weak projective measurement that are represented by $|\langle \omega \mid t \rangle|^2$ are quadratic in the measurement strength $\theta$ and can therefore be neglected in the weak limit. As the measurement strength increases, the contribution from this background increases, and its non-negligible contributions to the count rates should be taken into account in the evaluation of the experimental data.

Conveniently, the statistical background introduced by the projective measurement of time is approximately constant over the bandwidth considered. It is therefore possible to determine the background value experimentally, by simply selecting a post-selected frequency with vanishing signal contribution, $\hat{\rho}\mid \omega \rangle = 0$. At such post-selection frequencies, the coincidence rates are given by the background rate of
\begin{equation}
C_{\infty} = \frac{1}{8}\sin^2 \theta |\langle \omega \mid t \rangle|^2,
\end{equation}
where $|\langle \omega \mid t \rangle|^2 \approx 1/(2\pi)$ is independent of the specific frequency. In practice, it is usually possible to observe this background level at frequencies that lie well outside the spectrum of $\hat{\rho}$, but well within the much broader bandwidth of the reference pulse. With this background value, it is again possible to obtain the weak values of the projections directly from the count rates obtained at orthogonal polarizations of the probe photon outputs, where the background rates caused by the measurement back-action are subtracted from the values of the total counts,
\begin{eqnarray}
\label{eq:backsub}
\lefteqn{\hspace*{-0.8cm}
\frac{C_{\phi=\pi}-C_{\phi=0}}{(C_{\phi=\pi}+C_{\phi=0})- 2 C_{\infty}} =}
\nonumber \\ &&
 2 \sin \theta \cos \theta \;\; \mbox{Re}\left(\frac{\langle f \mid \Phi_{\mathrm{ref.}}\rangle \langle\Phi_{\mathrm{ref.}} \mid \hat{\rho} \mid f \rangle}{\langle f \mid \hat{\rho} \mid f \rangle} \right).
\end{eqnarray}
This background subtraction is particularly important when the post-selected frequency has a very low probability density in the input state, so that the background noise is much larger than the actual signal. By subtracting the background, it is possible to identify the correct value of the denominator in Eq.(\ref{eq:backsub}), which should be proportional to the actual spectrum of $\hat{\rho}$ given by $\langle \omega \mid \hat{\rho} \mid \omega \rangle$. Thus background subtraction provides a simple experimental method for correcting the errors introduced by measurement back-action, resulting in the same measurement results for all measurement strengths between $\theta=0$ and $\theta=\pi/4$. 

Since time and frequency are complementary, the post-selection of frequency results is particularly sensitive to the phases of superpositions between different input times. In particular, it is comparatively easy to obtain negative weak values for the projection operators on time by considering a coherent superpositions of different pulse times as a quasi-discrete input. By arranging the pulse times appropriately, the measurement of frequency can post-select an equal superposition of the initial pulse times with any combination of phase factors. It is then possible to perform temporal versions of observations of quantum paradoxes by weak measurements \cite{Val10,Res04,Lun09,Yok09}. Specifically, weak measurements of time - or even strong measurements of time, once the additional noise is subtracted - can show negative conditional probabilities that express the paradoxical features of quantum statistics in terms of non-classical correlations between photon energy and photon arrival times. The weak values of the projector on time determined in the present measurement therefore provide an experimentally accessible characterization of non-classical correlations in the time domain, illustrating the fact that quantum paradoxes observed in the spatial or polarization degrees of freedom apply equally to the physics of photon propagation. 

The complete access to the quantum features of photon arrival time and photon energy achieved by the present measurement setup can also be used to fully characterize the time-energy states of photons. Specifically, it has already been pointed out that the combination of weak projective measurements in one basis combined with a final measurement in a complementary basis provide a complete description of the quantum state, where the complex phases of the weak values describe the quantum coherence of the state \cite{Lun11,Lun12,Hof12a,Hof12b}. In the present case, the phase sensitive part of the count rate $C(\omega,t;\phi)$ in Eq.(\ref{eq:result}) corresponds to the Kirkwood function representation of the density operator $\hat{\rho}$ \cite{Kir33}, from which the Hilbert space expression of the operator can be reconstructed by
\begin{equation}
\hat{\rho} = \int \left(\langle \omega \mid t \rangle \langle t \mid \hat{\rho}\mid \omega \rangle \right) \frac{\mid t \rangle \langle \omega \mid}{\langle \omega \mid t \rangle} \; dt \; d\omega.
\end{equation}
It is therefore possible to perform a complete characterization of the temporal quantum state $\hat{\rho}$ by observing the count rate fringes of the phase $\phi$ that represents the ellipticity of the detected probe photon polarization, 
\begin{eqnarray}
\lefteqn{\frac{1}{2 \pi} \int_0^{2 \pi} \mathrm{e}^{-i \phi} C(\omega,t;\phi) d \phi =}
\nonumber \\ && \hspace*{1cm}
-\frac{1}{8}\sin\theta \cos\theta \;\; \langle \omega \mid t \rangle \langle t \mid \hat{\rho}\mid \omega \rangle.
\end{eqnarray}
Since the Kirkwood function is normalized, it is not even necessary to use the precise value of $\theta$ in the evaluation of the experimental data. Instead, it is sufficient to divide the unnormalized result obtained directly from the coincidence counts by their integral over time and frequency. The sequential measurement of time and frequency realized by two-photon interference at a polarization beam splitter thus represents a particularly direct method of quantum state tomography, where the data obtained corresponds directly to a fundamental phase space representation of the quantum state. 

\section{Wavefunction measurement and evaluation of entanglement}

For pure states, the Kirkwood function consists of a product of the wavefunctions in time and in frequency, modified only by the state independent factor $\langle \omega \mid t \rangle$. The $t$-dependence of the Kirkwood function obtained for a single post-selected frequency therefore corresponds to the wavefunction of the input state, so that a measurement of the Kirkwood function for a single frequency can be interpreted as a direct measurement of the quantum wavefunction \cite{Lun11}. We can now apply this insight to the coincidence count rates obtained by post-selecting only a single output frequency $\omega$ and scanning the peak time of the reference pulse. The normalized wavefunction is then given by
\begin{equation}
\psi(t) = \frac{1}{\sqrt{N_t}} \int_0^{2\pi} \mathrm{e}^{-i(\omega t + \phi)} C(\omega,t;\phi)d \phi,
\end{equation}　
where $N_t$ is the normalization factor obtained by integrating the squared result over time. This result shows that the present measurement setup can be used to directly measure the temporal wavefunction at arbitrary measurement strengths. In addition, it is possible to exploit the symmetry of time and frequency in the Kirkwood function to measure the wavefunction in the frequency basis by selecting a single reference pulse time $t$ and scanning the different output frequencies instead. The wavefunction can then be obtained from the frequency dependence of the count rate by
\begin{equation}
\psi(\omega) = \frac{1}{\sqrt{N_\omega}} \int_0^{2\pi} \mathrm{e}^{i(\omega t + \phi)} C(\omega,t;\phi)d \phi,
\end{equation}
where $N_\omega$ is the normalization factor obtained by integrating the squared result over all frequencies. It may be worth noting that dispersion effects appear as phase shifts in the frequency representation of the photon state, so that the above result can provide a prescription for the active compensation of unintended dispersion effects. In particular, it is possible to compress the pulse, corresponding to an optimization of the overlap between the photon state and the short-time pulse at $t$ by a unitary transform conserving the frequency $\omega$ \cite{Hof11}.

Since the measurement strength is not restricted to the weak limit, it is possible to apply the method to multi-photon states by interacting each photon with a separate reference pulse and measuring the coincidence count rates for all of the photons. For entangled photon pairs, the two photon wavefunction in time can then be obtained from the coincidence count rates post-selected at frequencies $\omega_1$ and $\omega_2$ while varying the local output polarizations $\phi_1$ and $\phi_2$. The complete entangled state can be obtained from
\begin{eqnarray}
\psi_2(t_1,t_2) &=& 
\nonumber \\ 
 &&  \hspace*{-1cm}\frac{1}{\sqrt{N_{2t}}} \int_0^{2\pi}\int_0^{2 \pi} \mathrm{e}^{-i(\omega_1 t_1 + \omega_2 t_2 + \phi_1 + \phi_2)} 
\nonumber \\ 
&\times& \;
C(\omega_1,t_1;\omega_2,t_2;\phi_1,\phi_2) \; d \phi_1 d\phi_2,
\end{eqnarray}　
where $N_{2t}$ is the normalization of the two-photon wavefunction. Thus, it is possible to directly measure the two-photon wavefunction of an entangled pair. In combination with sufficiently short reference pulses, the present method could thus be applied to obtain a direct and complete characterization of the time-energy entanglement of photon pairs generated in broad-band down-conversion or similar quantum optical processes. 

\vspace{0.1cm}

\section{Conclusions}

We have shown how a weak projective measurement of time can be realized by controllable two-photon interference at a polarizing beam splitter. A complete analysis of the measurement for all interaction strengths shows that it is also possible to obtain the weak values of the measurement projection at higher measurement strengths. The present setup therefore simplifies the experimental requirements and allows measurements with much better signal-to-noise ratios than conventional weak measurements. In combination with a final measurement of frequency, the projective measurement of photon arrival time provides a complete characterization of temporal and spectral coherence in the quantum state. In general, a complete description of the quantum state in energy and time is obtained in the form of the complex-valued Kirkwood function. For pure states, the measurement data obtained at fixed output frequency corresponds to the temporal wavefunction, and the data obtained at fixed time corresponds to the spectral wavefunction. Since the relevant weak value data can be obtained at any measurement strength, the method can also be applied to multi-photon states. In particular, the two-photon wavefunctions of time-energy entangled photon pairs can be obtained from the correlated four-photon coincidences obtained by performing separate measurements on the two entangled photons. The experimental procedure proposed here could therefore be extremely helpful in opening up new possibilities for quantum information processes with time-energy entangled photons. 

This work was supported by JSPS KAKENHI Grant Number 24540427.

\end{document}